%% file: main_evorobust_camera_ready.tex
\documentclass{article}

\usepackage[dblblindworkshop,final]{neurips_2026}
\workshoptitle{Self-Evolving Diversity-Driven Search for Robust AI Systems}

\usepackage[utf8]{inputenc}
\usepackage[T1]{fontenc}
\usepackage{url}
\usepackage{booktabs}
\usepackage{amsfonts}
\usepackage{amsmath}
\usepackage{microtype}
\usepackage{graphicx}
\graphicspath{{figures/}}
\usepackage{float}
\usepackage{color}
\definecolor{projectred}{rgb}{0.65,0.1,0.1}
\usepackage{hyperref}
\hypersetup{colorlinks=true, allcolors=black}
\newif\ifshort
\shorttrue

\title{Impossible Code, Barely Changed Score:\\
A Replay Study of Judge-Based Evaluation}

\author{%
  Masoud Fetanat \\
  UNSW Sydney \\
  \texttt{m.fetanat@unsw.edu.au} \\
}

\begin{document}

\maketitle

\begin{abstract}
Language models are often used to judge generated code without running it.
A useful evaluation should reveal substantial changes in the answers it
scores. We examine this requirement by repeating an earlier evaluation on
$421$ coding problems, comparing \texttt{Qwen-2.5-7B-Instruct} with two
versions trained using Group Relative Policy Optimization (GRPO).
One receives rewards from execution tests (CE-trained), and the other from
a strong language-model judge (SLJ-trained). The base model's syntax-failure
rate remains stable, but the trained checkpoints produce invalid Python on
$63\%$--$67\%$ of attempts. Despite this change, the strong
judge's Pass@$32$ scores fall by only about $3$ percentage points, from
$96.67\%$ to $93.95\%$ for SLJ-trained. Within the later run, allowing more
attempts makes the base and trained checkpoints look increasingly similar
under the judge. With $128$ attempts per question, the judge accepts at least one
SLJ-trained answer on $96.44\%$ of problems, compared with $28.27\%$ under
execution tests. We recommend reporting the share of answers that pass Python's
syntax check alongside every judge-based score. This check requires no
tests or code execution and flags invalid answers that the judge accepts,
although it cannot establish functional correctness. Our findings show how
a high judge-based score can hide substantial output damage and motivate
checking an evaluation's sensitivity to change.
Project page and code:
\begingroup
\hypersetup{urlcolor=projectred}
\url{https://curiousorbitnotes.github.io/evorobust-replay-audit/}\endgroup.
\end{abstract}

\section{Introduction}
\label{sec:intro}

Reinforcement learning with verifiable rewards (RLVR) trains language models
using automated feedback on their answers
\citep{lambert2024tulu3,guo2025deepseekr1}. For coding tasks, running programs
against tests provides direct evidence of correctness
\citep{chen2021humaneval}, although the result depends on the tests
\citep{liu2023evalplus}. Execution takes time and requires a sandbox, so
many pipelines instead ask a language model to read and grade the code
\citep{zheng2023judging,tong2024codejudge}. Such judgments can also provide
rewards for further training \citep{yuan2024selfrewarding}.

This choice introduces weaknesses that a high score may conceal.
Code-judge benchmarks already examine syntax errors and other failures
\citep{zhao2025codejudgeeval,jiang2025codejudgebench,wang2025axiom}.
Judges can favour familiar-looking answers \citep{panickssery2024selfpref},
and optimising a learned scorer can increase its score while true quality
falls \citep{gao2023overoptimization}. Models may also exploit a checker
\citep{helff2026gamingverifiers} or appear to improve under rewards with no
task information \citep{shao2025spuriousrewards}. Constructed benchmarks
control the mixture of inputs, but the answers produced by a deployed model
can change. We examine whether judge-based scores reveal such a change.

This question also matters when an evaluation is repeated.
Reported performance can vary with the evaluation code, decoding settings
and random seed \citep{hochlehnert2025soberlook}. Even greedy decoding can
change with hardware and numerical precision \citep{yuan2025nondeterminism},
and measurement gaps can inflate apparent RLVR gains
\citep{wu2025hiddencosts}. These findings motivate checking how results
change across runs. Our study examines a complementary risk: the reported
score may remain high even when the generated answers deteriorate.

We revisit a coding evaluation of models trained with GRPO
\citep{shao2024grpo}. We compare the base model with two trained checkpoints,
one rewarded by execution tests and the other by a strong AI judge.
When we repeated generation using the released checkpoints, the base model's
syntax-failure rate stayed similar, but the trained checkpoints produced many
invalid answers while retaining high judge-based scores. A simple syntax
check exposed the change. Missing records of the original model files and
serving setup prevent us from identifying its cause, and our findings do not
establish reward gaming or GRPO as responsible.
Sections~\ref{sec:setup} and~\ref{sec:judge} give the setup and detailed results.

The broader concern is the reliability of automated evaluation loops.
Program-search systems such as FunSearch \citep{romera2024funsearch} and
AlphaEvolve \citep{novikov2025alphaevolve} evaluate candidates by running them,
while self-rewarding training can rely on model judgments
\citep{yuan2024selfrewarding}. In a loop that selects candidates using a judge,
mistaken approvals can influence what is retained. Our study therefore
treats sensitivity to changes in generated answers as a property that should
be checked alongside individual judgments. Reporting a cheap syntax check
beside the judge's score would have made this incident visible.

\ifshort\else
\paragraph{Contributions.} What is new is not that judges mishandle
uncompilable code, but what that costs when nobody is checking. We quantify one
such failure in situ, where one checkpoint's validity fell from $99.58\%$ to
$33.10\%$ between two passes while the same judge's reported Pass@$32$ fell by
only $2.73$ pp. We also show how a
weak preference for code that parses ($83.3\%$ acceptance against $60.6\%$)
becomes an aggregate failure once most outputs are impossible, leaving $59.5\%$
of positive verdicts on programs that cannot run. We then evaluate a
necessary-condition gate that separates the two passes by two orders of
magnitude and is sound on all $70{,}239$ invalid samples.
\fi

\section{Methodology}
\label{sec:setup}

We test three code-writing versions of \texttt{Qwen-2.5-7B-Instruct}
\citep{qwen2024qwen25}: the \emph{base} model before additional GRPO training,
CE-trained (GRPO rewarded by execution tests), and SLJ-trained (GRPO
rewarded by a strong AI judge). These models generate the answers that the judges below evaluate. We use $421$ coding
questions: $164$ from HumanEval \citep{chen2021humaneval} and $257$ from the sanitized MBPP test split
\citep{austin2021mbpp}.

The \emph{first run} saved $32$ answers per question during the original
study ($13{,}472$ per checkpoint). The \emph{second run} generated $128$
($53{,}888$ per checkpoint) from the released checkpoints, using the same
questions on different hardware. The base model helps
check for changes common to every replayed model.

A \emph{sample} is one generated answer. The study's extractor, also used
by its execution reward, selects the code. Empty code or rejection by
Python's \texttt{ast.parse} makes a sample \emph{invalid}. Passing the check does not establish correctness. Judges read the question, reference and full answer
without running the code. Long inputs may be truncated. The parser sees only the
extracted code. We use \texttt{Qwen-2.5-72B-Instruct} as the strong judge and \texttt{Qwen-2.5-7B-Instruct} as the weaker judge. Both use vLLM \citep{kwon2023vllm} at
temperature $0$ and issue one correct/incorrect verdict per answer.

Execution runs the full HumanEval prompt and \texttt{check} function, or the
MBPP preamble and assertions. A pass requires a success token after testing. A normal exit without that token fails. This guards against false passes, but is not a security
boundary against adversarial code (Appendix~\ref{app:hardened}). Saved judge verdicts are assumed to match
their answers. A recheck supports this for the Qwen base model's second-run answers, but matching remains unverified for the trained runs (Appendix~\ref{app:judges}).

The \emph{syntax gate} requires judge approval and passing the syntax check.
Pass@$k$ estimates the chance of at least one success in $k$ attempts per
question under the chosen scorer, using the unbiased estimator of
\citet{chen2021humaneval}. For example, one judge-approved answer among
$32$ attempts is enough for success under the judge at that budget, even
if the other $31$ answers fail. A high score therefore does not imply that
most individual answers pass the check. Scores and rates are
percentages, and differences are measured in percentage points (pp). For within-run
differences, $95\%$ intervals resample the $421$ questions $5000$ times,
keeping each question's compared scores together (paired bootstrap).
Wider intervals indicate less precise estimates of the score differences.

\section{Results}

\subsection{Repeated evaluation}
\label{sec:audit}

\begin{figure}[t]
  \centering
  \includegraphics[width=\ifshort0.9\else1.0\fi\linewidth]{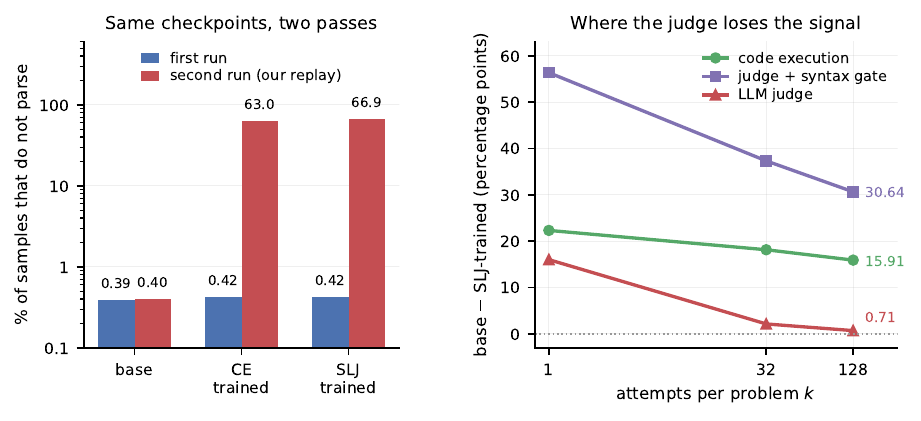}
  \caption{\textbf{Left:} how often each model wrote code that does not parse, in the first run and in the second. The base model's rate remains nearly the same. Both trained checkpoints jump from under $1\%$ to over $60\%$. The scale is logarithmic, so the short bars are visible. \textbf{Right:} how far apart the base model and the damaged SLJ-trained checkpoint look, measured as a score difference in percentage points (pp) on the same second-run outputs, as the number of attempts allowed per problem grows. Running the code keeps them clearly apart. The judge does not: its gap has almost vanished by $128$ attempts. Scores and gaps are listed in Appendix~\ref{app:figurevalues}.}
  \label{fig:replay}
\end{figure}

Table~\ref{tab:validity} reports invalid answers. The base changes little:
$0.39\%$ in the first run and $0.40\%$ in the second.
CE-trained rises from $0.42\%$ to $63.04\%$, and SLJ-trained from $0.42\%$ to
$66.90\%$. Both benchmarks show the increase. For SLJ-trained, more than
half the $128$ answers are invalid on
$283$ of $421$ questions, and all answers pass the syntax check on only $29$.
For the base model, these counts are $0$ and $363$, respectively
(Appendix~\ref{app:taxonomy}).

\begin{table}[t]
\centering
\caption{Percentage of answers with empty code or invalid Python syntax.
First run: $32$ answers per question, saved in the original study. Second:
$128$ new answers from the released checkpoints, on the same $421$ questions
but different hardware. Incomplete records prevent checking
whether the model files and serving settings matched exactly
(Appendix~\ref{app:settings}). HumanEval/MBPP gives the rate for each benchmark.}
\label{tab:validity}
\small
\begin{tabular}{lcccc}
\toprule
& \multicolumn{2}{c}{first run} & \multicolumn{2}{c}{second run} \\
\cmidrule(lr){2-3}\cmidrule(lr){4-5}
checkpoint & invalid & HEval / MBPP & invalid & HEval / MBPP \\
\midrule
\texttt{Qwen-2.5-7B} base & $0.39\%$ & $0.3$ / $0.4\%$ & $0.40\%$ & $0.1$ / $0.6\%$ \\
CE-trained                     & $0.42\%$ & $0.2$ / $0.5\%$ & $63.04\%$ & $64.8$ / $61.9\%$ \\
SLJ-trained                    & $0.42\%$ & $0.1$ / $0.6\%$ & $66.90\%$ & $68.4$ / $66.0\%$ \\
\bottomrule
\end{tabular}
\end{table}

Most failures involve a closing bracket with no matching opening bracket,
or with a different bracket type: $35{,}562$ of $36{,}051$
SLJ-trained invalid samples ($98.6\%$). For example,
\mbox{\texttt{for i in range(1, len(numbers))):}} has an extra closing parenthesis.
Early cutoff would tend to leave unfinished code, such as an unclosed opening
bracket, near the length limit. Only one of the $36{,}051$ invalid samples
has that pattern, so early cutoff does not
explain the dominant error
(Appendix~\ref{app:truncation}).

\ifshort\else
Extraction is not the explanation either. Taking the last fenced block, as the
study's extractor does, leaves SLJ-trained at a $33.10\%$ parse rate. Accepting
the sample if \emph{any} fenced block parses gives $33.12\%$
(Appendix~\ref{app:extraction}).
\fi

\subsection{Scorer comparison across sampling budgets}
\label{sec:judge}

Table~\ref{tab:temporal} in Appendix~\ref{app:additionalstats} compares the
two runs at Pass@$32$, using the Qwen-2.5-72B-Instruct judge.
SLJ-trained scores $96.67\%$ in the first run and $93.95\%$ in the second,
a drop of $2.73$ pp. CE-trained shows a similar decrease, while the base
model's score changes little. The judge scored the later answers in a
separate run, so changes in its scoring could have contributed to these
decreases. The base model's small measured change does not establish that
the judge stayed stable. Its uncertainty interval still allows score
changes of a few percentage points (Table~\ref{tab:temporal}). We therefore
cannot separate the effects of changed answers and changed judging in
this comparison between runs.

Meanwhile, SLJ-trained's invalid-answer rate rises from $0.42\%$ to
$66.90\%$, a change of $66.5$ pp (Table~\ref{tab:validity}).
The judge score drops by less than three points while most answers become
invalid. These measurements describe different things: Pass@$32$ concerns
at least one approved answer per question, while invalidity counts individual
answers. We therefore do not divide one change by the other.

\ifshort\else
In absolute terms the judge's fall is equivalent to about $11$ of $421$
problems, against roughly $35{,}800$ of $53{,}888$ samples ceasing to be
Python. Per-question verdicts were not retained for the first run, so this is
a difference of rates rather than a count of problems that changed.
\fi

\ifshort\else

\begin{table}[t]
\centering
\caption{Every scorer, at every attempt budget, on the second-run outputs. Scores are percentages, and the bottom row gives differences in pp. The bottom row is what matters: how far each scorer places the base model above the damaged one. Running the code keeps them $22$ pp apart at one attempt and $16$ pp apart at $128$. The judge starts at $16$ pp and ends at $0.7$ pp. Only the judge loses the signal, and it loses it at the budgets papers report.}
\label{tab:kgrid}
\footnotesize
\setlength{\tabcolsep}{2.5pt}
\begin{tabular}{lccc|ccc|ccc}
\toprule
& \multicolumn{3}{c|}{judge} & \multicolumn{3}{c|}{judge $+$ gate} & \multicolumn{3}{c}{execution} \\
checkpoint & $k{=}1$ & $32$ & $128$ & $k{=}1$ & $32$ & $128$ & $k{=}1$ & $32$ & $128$ \\
\midrule
\texttt{Qwen-2.5-7B} base & $84.12\%$ & $96.10\%$ & $97.15\%$ & $83.92\%$ & $96.05\%$ & $97.15\%$ & $33.74\%$ & $42.88\%$ & $44.18\%$ \\
CE-trained                & $67.45\%$ & $92.70\%$ & $95.25\%$ & $30.05\%$ & $59.23\%$ & $66.27\%$ & $12.39\%$ & $26.32\%$ & $30.17\%$ \\
SLJ-trained               & $68.10\%$ & $93.95\%$ & $96.44\%$ & $27.59\%$ & $58.74\%$ & $66.51\%$ & $11.42\%$ & $24.73\%$ & $28.27\%$ \\
\midrule
base $-$ SLJ-trained (pp) & $16.01$ & $2.15$ & $0.71$ & $56.34$ & $37.31$ & $30.64$ & $22.32$ & $18.15$ & $15.91$ \\
\bottomrule
\end{tabular}
\end{table}
\fi

Table~\ref{tab:judgefull} in Appendix~\ref{app:judge} shows that the judge
often approves invalid code. It accepts $60.56\%$ of SLJ-trained answers
that fail the syntax check, compared with $83.35\%$ of those that pass.
Of all answers it approves for this checkpoint, $59.49\%$ are invalid.
In other words, more than half of its approvals are for code that cannot
run. CE-trained shows the same pattern. The judge favours answers that
parse, but does not require valid syntax before approving them. Such
mistakes have little effect when invalid answers are rare, as in the base
run. They affect many more approvals when most generated answers are
invalid. These results describe the second-run answers and do not establish
that the judge's error rates worsened between runs.

\begin{table}[t]
\centering
\caption{Four checks on the same second-run answers. Parsing measures the
percentage of individual answers that pass Python's syntax check. The other
columns report Pass@$128$: the estimated chance of at least one success in
$128$ attempts per question. Success means judge approval, approval plus
passing the syntax check (gate), or passing execution tests, respectively.
Judge scores are similarly high for all three models, whereas the other checks distinguish the base from the trained models. CE-trained and SLJ-trained
remain close.}
\label{tab:instruments}
\small
\begin{tabular}{lcccc}
\toprule
checkpoint & \% samples parsing & judge Pass@$128$ & judge $+$ gate & execution Pass@$128$ \\
\midrule
\texttt{Qwen-2.5-7B} base & $99.60\%$ & $97.15\%$ & $97.15\%$ & $44.18\%$ \\
CE-trained                     & $36.96\%$ & $95.25\%$ & $66.27\%$ & $30.17\%$ \\
SLJ-trained                    & $33.10\%$ & $96.44\%$ & $66.51\%$ & $28.27\%$ \\
\bottomrule
\end{tabular}
\end{table}

Table~\ref{tab:instruments} compares the same second-run answers under
each check. Between the base and SLJ-trained checkpoints, the judge's
Pass@$128$ differs by only $0.71$ pp, while execution differs by $15.91$ pp
and the share of answers that parse differs by $66.50$ pp.

Table~\ref{tab:figurevalues} in Appendix~\ref{app:figurevalues} lists the
scores at each attempt budget. Table~\ref{tab:paired} gives the gaps and their
$95\%$ intervals. The judge's gap between the base and SLJ-trained
checkpoints shrinks from $16.01$ pp at one attempt to $0.71$ pp at $128$.
Execution keeps a larger gap, decreasing from $22.32$ pp to $15.91$ pp.
All these scores use the same second-run answers. Changing $k$ changes the
number of attempts allowed in the score calculation, without generating
new answers. This asks whether each scorer can distinguish the two
checkpoints. The earlier-versus-later comparison is available only at
$k=32$.

More attempts give each question more chances to receive an approval.
Even when individual answers are unreliable, the chance of finding at
least one approved answer can be high. Here, about a third of SLJ-trained
answers still parse, and the judge also approves many invalid answers.
Both contribute to its high multiple-attempt score. Its Pass@$128$ is
$96.44\%$, compared with $66.51\%$ after removing approvals
for invalid code (Table~\ref{tab:instruments}). The high multiple-attempt
score therefore hides much of the difference in the generated answers.

\ifshort\else
Judge leniency is a separate matter. Even on the base model the judge
reads $97.15\%$ where execution reads $44.18\%$, and that gap is a known problem
we do not claim here.
\fi

\subsection{The syntax gate}
\label{sec:gate}

Table~\ref{tab:gatecost} in Appendix~\ref{app:additionalstats} shows how much
judge-based Pass@$1$ falls when approvals for invalid code are removed.
The drop is $0.19$ pp for the base model, $37.40$ pp for CE-trained and $40.51$ pp for
SLJ-trained. Table~\ref{tab:gatecost} gives the paired $95\%$ intervals
for all three drops. The answers stay the same throughout this comparison.
The gate adds one requirement to the judge's approval: the extracted code
must also pass Python's syntax check. It therefore measures how much the
reported score relies on approvals for invalid code.

The gate necessarily removes approvals for invalid code. What matters is
how many it removes: the score drop is more than a hundred times larger
for the damaged runs. In this incident, any invalidity threshold between
$1\%$ and $50\%$ would distinguish the healthy and damaged runs
(Table~\ref{tab:validity}). An earlier
run can provide a reference value. Compiler and parser checks already serve
as training rewards \citep{le2022coderl,shojaee2023ppocoder}. Our recommendation
is to report such a check alongside judge-based scores.

None of the $70{,}239$ invalid answers passed the execution tests
(Table~\ref{tab:gatecost}). However, code that passes the syntax check can
still be wrong. SLJ-trained's gated Pass@$128$ is $66.51\%$, while execution
gives $28.27\%$ (Table~\ref{tab:instruments}). The gate flags damaged outputs
without establishing correctness or explaining the cause.

\ifshort\else
The general point is about where a guarantee comes from. A parser decides
membership in a language and its verdict holds whatever the model, prompt or
sampling budget. A judge returns an opinion whose reliability is itself an
empirical question, and whose sensitivity we have just seen vary by $22\times$
with the sampling budget alone. Pairing a cheap sound check with an expensive
unsound one is not a new idea in verification. This incident shows the cost of
omitting the cheap half.
\fi

\section{Discussion and limitations}
\label{sec:limitations}

Both damaged checkpoints share a release pipeline, so this is closer to
one incident than two independent failures. We do not estimate how often
such failures occur. The base model loads from a public model identifier,
while the trained checkpoints load from local directories
(Appendix~\ref{app:settings}). Its stable outputs argue against a problem
affecting every model, but do not rule out a fault in loading trained checkpoints.

We have not tested how an alarm threshold responds to harmless changes in
the generation setup. In a controlled check, adding a surplus bracket to about
$10\%$ to $60\%$ of base-model answers changes judge Pass@$32$ by at most
$0.24$ pp, while the parse rate falls to between $89.3\%$ and $39.9\%$
(Table~\ref{tab:corruption}, protocol in Appendix~\ref{app:signals}).
This supports using the parser to flag this type of damage, but does not
establish a general monitoring threshold. The cause of the damaged
checkpoint outputs remains unknown.

Execution scores also depend on the benchmark prompts. MBPP omits the
function name expected by its tests, so an answer using a different name
can fail. In a separate Qwen base-model run with function names supplied,
MBPP Pass@$128$ rises from $12.45\%$ to $79.38\%$
(Table~\ref{tab:mbppnames}, execution details in Appendix~\ref{app:hardened}).
Those results use new answers and do not replace the original evaluation.
Absolute execution scores and judge--execution gaps should be read with
this naming issue in mind. Our findings concern code, and comparisons
between checkpoints describe the specific second-run answers. Further checks
and settings appear in Appendices~\ref{app:scope}--\ref{app:additionalstats}.

To make changes between runs easier to investigate, each run should retain
the generated answers, the judge's inputs and verdicts, the generation
settings, and checksums of the model files. These records connect a reported
score to the outputs and setup that produced it. Without them, a later run
may expose a large discrepancy while leaving its cause unresolved, as
happened here.

\section{Conclusion}

\ifshort
A high judge-based score can coexist with extensive damage to generated
code. In this study, the trained checkpoints produced many answers that
Python could not parse, while their judge-based scores remained high.
Allowing more attempts made this damage harder to see in the strong
judge's score, because a question needs only one approved answer to succeed
at Pass@$k$. Execution tests continued to distinguish the base and damaged
checkpoints on the same outputs. These observations concern one replay
incident whose cause remains unresolved, but they support a practical
reporting recommendation. We recommend reporting the share of answers
that parse alongside every judge-based score and comparing that share
across runs. The syntax check does not establish functional correctness,
but it reveals whether a high score coexists with many invalid answers.
Reporting both measures therefore helps readers distinguish finding an
approved answer from consistently generating valid code, and notice output
changes that the headline score alone may hide.
\else

A judge-based score can sit close to where it was while the thing it scores
stops being a program. Here the reported Pass@$32$ moved by $2.73$ pp while the
share of outputs that were valid Python moved by $66.5$ pp, and by $k=128$
the judge's separation between a working checkpoint and a damaged one had
shrunk twenty-two-fold.

The recommendation is one extra column in a results table. Beside every
judge-based number, report the share of outputs that satisfy a cheap necessary
condition, which for code is whether the output parses at all. It costs a
healthy run $0.19$ pp Pass@$1$ and flags a damaged one by $40.51$ pp, it needs no
tests and no sandbox, and it is available to anyone who re-runs an evaluation
on new hardware or a new serving stack.

The wider point is the one this paper opened with. A judge can be accurate on
the inputs it was tested against and still fail to move when those inputs
change character, because accuracy and sensitivity are different properties and
only the first is usually measured. Nothing here argues against using judges.
They are used because execution is expensive, and that remains true. What this
incident argues for is reporting, beside the judgment, one thing the judgment
cannot fake. For code that thing is the parser. Other domains admit their own
cheap necessary conditions, a well-formed final answer or a parseable
expression, and whether each is as cheap as this one is a question we have not
settled.
\fi

\begin{ack}
This research includes computations using the computational cluster Katana
supported by Research Technology Services at UNSW Sydney
(\href{https://doi.org/10.26190/669x-a286}{doi:10.26190/669x-a286}). No
external funding supported this work.
\end{ack}

\clearpage
\bibliographystyle{plainnat}
\bibliography{references}

\input{appendix_camera_ready}

\end{document}

%% file: appendix_camera_ready.tex
\appendix

\section{Failure taxonomy}
\label{app:taxonomy}

Table~\ref{tab:taxonomy} lists the parse errors raised by \texttt{ast.parse} on
the second-run outputs, grouped by message with variable content collapsed.
Most errors fall into a small number of categories. For both
trained checkpoints a single class, a closing bracket with no matching opener,
covers more than $82\%$ of failures on its own, and the two bracket classes
together cover $98.4\%$ and $98.6\%$. On the base model the same two
classes account for $12.0\%$ of its $216$ failures, which are spread across
$16$ distinct parser messages and two empty outputs.

\begin{table}[h]
\centering
\caption{What the Python parser complained about, grouped by message. Counts are individual samples out of $53{,}888$ per checkpoint. The bracket rows include the f-string forms of the same messages, and empty outputs count under other classes. Both trained checkpoints mainly fail through surplus closing brackets. The base model has few failures, spread across many causes.}
\label{tab:taxonomy}
\small
\begin{tabular}{lrrr}
\toprule
parse error & base & CE-trained & SLJ-trained \\
\midrule
unmatched closing bracket             & $20$  & $28{,}115$ & $29{,}924$ \\
closing bracket does not match opener & $6$   & $5{,}305$  & $5{,}638$  \\
invalid syntax (unclassified)         & $153$ & $250$      & $192$      \\
positional argument follows keyword   & $0$   & $114$      & $118$      \\
other classes                         & $37$  & $188$      & $179$      \\
\midrule
total failing                         & $216$ & $33{,}972$ & $36{,}051$ \\
\bottomrule
\end{tabular}
\end{table}

The errors are easy to see in the generated code. The following examples show
the last line of two extracted code blocks from \texttt{slj\_code\_s1}:

\begin{quote}
\texttt{for i in range(1, len(numbers))):}
\end{quote}

\begin{quote}
\texttt{print(has\_close\_elements([1.0, 2.0, 3.0]], 0.5)))}
\end{quote}

In both the surrounding code looks plausible and is correctly indented. The second
carries two surplus closing tokens rather than one.

\paragraph{Failures across questions.} For each question, invalidity is the
percentage of its $128$ answers that fail the parse check. Sorting these
rates for SLJ-trained and reading them at $10\%$ intervals from the minimum
to the maximum gives $0\%$, $3\%$, $27\%$, $45\%$, $68\%$, $84\%$,
$94\%$, $98\%$, $100\%$, $100\%$, $100\%$: $283$ of $421$ prompts are more than
half invalid and $105$ are entirely invalid, with only $29$ having no invalid answers. The worst
$10\%$ of prompts contribute just $14.9\%$ of all invalid samples, so no small
subset drives the aggregate. The base model's failures are less frequent and more concentrated: no
prompt exceeds $50\%$ invalid, $363$ of $421$ are clean, and its worst $10\%$
of prompts account for $92.6\%$ of its invalid answers. This pattern is
consistent with infrequent failures affecting relatively few questions.

\section{Truncation as an explanation}
\label{app:truncation}

A generation cut off at a length limit can leave unfinished code, such as
an \emph{unclosed} bracket at the end of input. Such failures may cluster
near the length limit. These signatures do not explain the dominant error.

Classifying every \texttt{slj\_code\_s1} parse failure by signature gives
$35{,}562$ with an extra closing bracket ($98.6\%$ of $36{,}051$), $484$ with
other syntax errors, $4$ empty, and $1$ with the truncation signature. Failing
samples have a median length of $298$ characters against $181$ for parsing
ones. We measured characters rather
than tokens, so this supports the bracket-signature argument rather than
independently locating the generation limit.

\section{Robustness to code extraction}
\label{app:extraction}

The reported rates use the study's own extraction function, the one its
execution reward calls, which takes the last fenced block of a response. We also checked two alternatives for \texttt{slj\_code\_s1}: accepting the
sample if \emph{any} fenced block parses gives $33.12\%$ valid against
$33.10\%$, and parsing the unextracted response gives $0.04\%$, which reflects
response formatting, including code fences and surrounding text. The base model
sits at $99.60\%$ under both fenced variants. The choice of code extraction therefore does not explain the low parse rate.

\section{Judge behaviour in full}
\label{app:judge}

Table~\ref{tab:judgefull} gives the acceptance rates behind
Section~\ref{sec:judge}, separated by whether the scored sample parses.

\begin{table}[h]
\centering
\caption{Strong-judge approval rates, separated by syntax validity. The middle columns give its approval rate on invalid and valid samples. Valid here means non-empty code accepted by Python's parser, without establishing functional correctness. The final column gives the percentage of approved answers that are invalid.}
\label{tab:judgefull}
\small
\begin{tabular}{lrrrr}
\toprule
checkpoint & \multicolumn{2}{c}{invalid samples} & valid samples & positives invalid \\
\cmidrule(lr){2-3}
& $n$ & rate & rate & \\
\midrule
\texttt{Qwen-2.5-7B} base & $216$    & $47.69\%$ & $84.26\%$ & $0.23\%$  \\
CE-trained                     & $33{,}972$ & $59.32\%$ & $81.31\%$ & $55.44\%$ \\
SLJ-trained                    & $36{,}051$ & $60.56\%$ & $83.35\%$ & $59.49\%$ \\
\bottomrule
\end{tabular}
\end{table}

Appendix~\ref{app:judges} also compares the strong and weak judges on the
CE-trained and SLJ-trained answers from the second run.

\section{Judge and generation configuration}
\label{app:judgecfg}

Both judges are served with vLLM \citep{kwon2023vllm} at temperature $0$
(greedy decoding). A re-check
reproduced $99.69\%$ of cached verdicts on the Qwen base-model subset (Appendix~\ref{app:judges}). Each scoring call is a single
chat completion over one (problem statement, reference solution, model response) triple,
in that order, and returns a binary correct/incorrect verdict. If the triple exceeds the
client's configured input budget, the scoring code trims it. Token-based trimming reduces the problem statement first, then the response, and the reference last. If the tokenizer is unavailable, the code falls back to character-based trimming. The Qwen base-model recheck found no request over the server's context limit. We did not measure truncation for the damaged checkpoints. In the original runs, network failures and non-2xx responses scored $0$. A later strict mode aborts instead, and the Qwen recheck found one cached zero that re-judged as $1$. The
strong judge is \texttt{Qwen-2.5-72B-Instruct} and the weak judge
\texttt{Qwen-2.5-7B-Instruct}, the latter sharing a family with the policy.

Judges score the full response text unless truncated. The parser sees only the extracted code
block. Parsing here means that the extracted code is non-empty and accepted by Python's \texttt{ast.parse}. It does not imply that the code compiles, runs or is correct.

\section{Execution scores}
\label{app:exec}

Table~\ref{tab:execfull} reports how often the second-run answers pass the
execution tests. We use a hardened harness, meaning a test runner with
additional checks against false passes (Appendix~\ref{app:hardened}). These
scores describe the saved answers under this test setup. We cannot verify
that both runs used identical model files, so the scores do not establish
the capabilities of the original trained checkpoints.

\begin{table}[h]
\centering
\caption{Execution Pass@$k$ on the second-run answers. Values are percentages.}
\label{tab:execfull}
\small
\begin{tabular}{lccc}
\toprule
checkpoint & Pass@$1$ & Pass@$32$ & Pass@$128$ \\
\midrule
\texttt{Qwen-2.5-7B} base & $33.74\%$ & $42.88\%$ & $44.18\%$ \\
CE-trained                     & $12.39\%$ & $26.32\%$ & $30.17\%$ \\
SLJ-trained                    & $11.42\%$ & $24.73\%$ & $28.27\%$ \\
\bottomrule
\end{tabular}
\end{table}

On every one of the $70{,}239$ invalid samples, the execution
verifier also withheld credit. We report this as a consistency check rather
than as independent confirmation: the interpreter must parse a program before
running it, so a non-parsing sample cannot pass, and the agreement is largely a
consequence of that. We found no instance of a false execution pass among
these invalid samples.

\section{Reproduction}
\label{app:repro}

The analyses of cached results run on CPU without generating new model answers. They require the cached generations, judge verdicts and execution scores. The scripts cover the following steps:

\begin{itemize}
\itemsep2pt
\item the comparison of the two runs, producing validity rates, the failure taxonomy, judge
  acceptance and the gate intervals.
\item the truncation check, producing the signature classification and length
  distributions in Appendix~\ref{app:truncation}.
\item the extraction check, comparing the three renderings in
  Appendix~\ref{app:extraction}.
\item the instrument comparison, producing the judge and gate columns of Table~\ref{tab:instruments} (this script still reads the cached execution scores).
\item the hardened sample-level scores (canon4/canon5 outputs, archived with checksums) and \texttt{evorobust\_analyses.py}, which produce the hardened execution values and the intervals in the main text.
\item \texttt{wljdam\_analyze.py} (weak judge on the damaged checkpoints) and \texttt{corrdose\_build.py} with \texttt{corrdose\_analyze.py} (milder corruption).
\end{itemize}

Bootstrap intervals resample the $421$ prompts with replacement, $5000$ times,
under a fixed seed. Pass@$k$ uses the unbiased estimator of
\citet{chen2021humaneval}. The analysis scripts, the judge prompt and the verified summary tables are available at \url{https://github.com/curiousorbitnotes/evorobust-replay-audit}, with a project page at \url{https://curiousorbitnotes.github.io/evorobust-replay-audit/}. Raw generations, checkpoints and sample-level scores are not part of that release, so it provides the reported summary tables but does not permit independent recomputation without those inputs.

\paragraph{Which scorer produced which number.} All execution results in
the main text use the hardened harness of Appendix~\ref{app:hardened}.
The generation archive also contains scores from an earlier scorer (OLD).
That version could define the HumanEval test function without calling it,
and so record false passes. We do not use those scores here.
Table~\ref{tab:hardened} compares the study's final cached scorer
(cached-NEW, stored as the \texttt{ce\_FINAL} arrays) with the hardened
runner, on the same answers from all three Qwen checkpoints. Judge
verdicts are unaffected by the execution-scorer changes because they do not
run the code. Parse rates are computed directly from the generations.

\section{Scope of the claims}
\label{app:scope}

The evidence comes from one incident in one training and release pipeline. We do
not estimate how common such failures are, and we do not claim a cause. The following paragraphs distinguish the incident-specific findings from the comparisons of scoring methods.

\paragraph{Specific to this incident.} The two trained checkpoints, as we
downloaded them, wrote non-parsing Python on $63.0\%$ and $66.9\%$ of attempts,
mostly through one surplus closing bracket. The Qwen base model's syntax-failure
rate remained stable. We could not preserve artifact identity between
the two runs, so we make no claim about byte-identical weights, hardware, or
serving software. Possible causes include checkpoint corruption, loading or
chat-template drift, tokenizer or decoding differences, and other artifact
changes. The available records do not let us distinguish these possible causes.

\paragraph{About judges and sampling budgets.} Holding the second-run outputs
fixed, the judge's separation of the base model from a damaged checkpoint
shrinks from $16.01$ pp at one attempt to $0.71$ pp at $128$, while execution keeps
a separation of $15$ pp or more. This result does not depend on why the outputs
are damaged, but it rests on cross-model contrasts within the second run, not on
a paired before-and-after comparison across $k$. Parse rate is a necessary
condition only. A program that parses can still be wrong, and a high parse rate
says nothing about functional correctness.

\section{Execution scores under the hardened harness}
\label{app:hardened}

The execution numbers in the main text come from a test runner applied to
the same second-run samples. We call it hardened because it includes
additional checks against false passes. For HumanEval, it executes the full
benchmark prompt, the generated code, and the benchmark's \texttt{check}
function. For MBPP, it uses a preamble and the benchmark assertions. A pass
requires an explicit success signal produced after the tests complete.
Exiting normally without that signal counts as a failure. The result-writing
function is captured before the generated code runs, which reduces the risk
of a false pass if that code changes output functions. These safeguards are
not a security boundary against adversarial code. MBPP prompts omit the
function name, so strict MBPP execution also fails answers that use another
name. In a separate run, we added the expected function name to each MBPP prompt and generated $128$ new answers per question with the Qwen base model. Its MBPP Pass@$128$ is $79.38\%$, compared with $12.45\%$ under the original prompts (Table~\ref{tab:mbppnames}). These new answers do not replace the second-run outputs used elsewhere in the paper. The harness passes all $421$
reference solutions and fails every negative control we tried (programs that
return nothing, exit cleanly, or forge the pass token). The execution reward of
the original study scores the samples differently: its cached scorer cannot
decompose the asserts of eleven HumanEval problems and scores them zero by
design. Table~\ref{tab:hardened} gives both. Pooled Pass@$k$ is higher under the
hardened harness by $0.6$ pp to $2.6$ pp. Across all $70{,}239$ invalid
samples the hardened harness credits none, as did the cached scorer.
The MBPP Pass@$128$ values are identical. The two scorers also differ on a
small number of individual samples (for example, the cached-NEW comparator treats a
tuple and a list as equal, which the hardened native assertions do not). The gap between base and damaged checkpoint moves
from $20.95$ pp to $22.32$ pp at $k=1$ and from $14.96$ pp to $15.91$ pp at $k=128$, so
the conclusions are unchanged.

\begin{table}[h]
\centering
\caption{Execution Pass@$k$ on the second-run outputs: the original study's cached scorer, and the hardened harness (the values in Table~\ref{tab:execfull} and the main text).}
\label{tab:hardened}
\small
\begin{tabular}{lcccccc}
\toprule
& \multicolumn{3}{c}{study's cached scorer (cached-NEW)} & \multicolumn{3}{c}{hardened} \\
checkpoint & @$1$ & @$32$ & @$128$ & @$1$ & @$32$ & @$128$ \\
\midrule
\texttt{Qwen-2.5-7B} base & $31.78\%$ & $40.49\%$ & $41.57\%$ & $33.74\%$ & $42.88\%$ & $44.18\%$ \\
CE-trained                & $11.71\%$ & $24.99\%$ & $28.74\%$ & $12.39\%$ & $26.32\%$ & $30.17\%$ \\
SLJ-trained               & $10.83\%$ & $23.43\%$ & $26.60\%$ & $11.42\%$ & $24.73\%$ & $28.27\%$ \\
\bottomrule
\end{tabular}
\end{table}

\section{Values behind \texorpdfstring{Figure~\ref{fig:replay}}{Figure 1}}
\label{app:figurevalues}

Table~\ref{tab:figurevalues} shows the scores behind the right panel of
Figure~\ref{fig:replay}. The figure compares the base model with SLJ-trained. The table also includes CE-trained so readers can compare all three checkpoints.
For each question, Pass@$k$ counts success when at least one of $k$ attempts
passes the chosen scorer. The judge counts approval by \texttt{Qwen-2.5-72B-Instruct} as a success. With the gate, the approved code must also parse. Execution requires the program to pass the tests.
The same saved second-run answers are used throughout: increasing $k$ allows
more attempts in the score calculation and does not change the model.

\begin{table}[h]
\centering
\caption{Pass@$k$ scores for all three checkpoints on the same $421$
second-run problems. The base is \texttt{Qwen-2.5-7B-Instruct}.
Figure~\ref{fig:replay} plots the base-minus-SLJ-trained gaps.
Scores are percentages, and gaps are measured in percentage points (pp).
Gaps use unrounded scores, so they can differ by $0.01$ pp from subtracting
the displayed scores. Execution uses the hardened harness
(Table~\ref{tab:hardened}). Table~\ref{tab:paired} gives both sets of gaps and their
$95\%$ intervals to one decimal place.}
\label{tab:figurevalues}
\small
\begin{tabular}{rccccc}
\toprule
& \multicolumn{3}{c}{Score (\%)} & \multicolumn{2}{c}{Base $-$ checkpoint (pp)} \\
\cmidrule(lr){2-4}\cmidrule(lr){5-6}
Attempts $k$ & base & CE-trained & SLJ-trained & CE-trained & SLJ-trained \\
\midrule
\multicolumn{6}{l}{\textbf{LLM judge (Qwen-2.5-72B-Instruct)}} \\
1 & $84.12$ & $67.45$ & $68.10$ & $16.67$ & $16.01$ \\
32 & $96.10$ & $92.70$ & $93.95$ & $3.40$ & $2.15$ \\
128 & $97.15$ & $95.25$ & $96.44$ & $1.90$ & $0.71$ \\
\midrule
\multicolumn{6}{l}{\textbf{judge + syntax gate}} \\
1 & $83.92$ & $30.05$ & $27.59$ & $53.87$ & $56.34$ \\
32 & $96.05$ & $59.23$ & $58.74$ & $36.82$ & $37.31$ \\
128 & $97.15$ & $66.27$ & $66.51$ & $30.88$ & $30.64$ \\
\midrule
\multicolumn{6}{l}{\textbf{code execution (hardened)}} \\
1 & $33.74$ & $12.39$ & $11.42$ & $21.35$ & $22.32$ \\
32 & $42.88$ & $26.32$ & $24.73$ & $16.57$ & $18.15$ \\
128 & $44.18$ & $30.17$ & $28.27$ & $14.01$ & $15.91$ \\
\bottomrule
\end{tabular}
\end{table}

\section{Paired per-problem analysis with uncertainty}
\label{app:paired}

All instruments score the same $421$ problems, so differences are taken per problem and resampled over problems ($5{,}000$ bootstrap draws, $95\%$ intervals). Table~\ref{tab:paired} gives the separation between the base model and each trained checkpoint. These comparisons use different checkpoints within the second run.

For SLJ-trained, the judge's separation is $16.01$ pp at $k=1$ and shrinks until its interval includes zero at $k=128$ ($0.71$ pp, $[-1.19,\,2.61]$ pp). Execution under the hardened harness keeps a separation whose interval excludes zero at every budget ($15.91$ pp, $[12.11,\,19.95]$ pp at $k=128$), and the parse gate keeps a larger one ($30.64$ pp). 

The CE-trained checkpoint shows a similar pattern (judge $1.90$ pp, $[0.00,\,4.04]$ pp, and execution $14.01$ pp, $[10.21,\,17.81]$ pp at $k=128$). 

The gap between the judge and hardened execution on the same outputs is $52.97$ pp for the base model ($[47.98,\,57.96]$ pp), $65.08$ pp for the CE-trained checkpoint and $68.17$ pp for the SLJ-trained one ($[63.66,\,72.68]$ pp), at $k=128$.

\begin{table}[h]
\centering
\caption{Base minus each trained checkpoint's Pass@$k$ per instrument, in percentage points (pp), with $95\%$ prompt-bootstrap intervals (paired over the same $421$ problems). Each cell gives the gap followed by its interval in brackets. Positive gaps favour the base model.}
\label{tab:paired}
\small
\begin{tabular}{lccc}
\toprule
instrument & $k=1$ & $k=32$ & $k=128$ \\
\midrule
\multicolumn{4}{l}{\textbf{Base $-$ CE-trained}} \\
judge & $16.7\ [13.8,\,19.7]$ & $3.4\ [1.4,\,5.4]$ & $1.9\ [0.0,\,4.0]$ \\
judge $+$ gate & $53.9\ [50.0,\,57.7]$ & $36.8\ [32.5,\,41.3]$ & $30.9\ [26.1,\,35.6]$ \\
execution (hardened) & $21.3\ [17.7,\,25.1]$ & $16.6\ [13.0,\,20.3]$ & $14.0\ [10.2,\,17.8]$ \\
\midrule
\multicolumn{4}{l}{\textbf{Base $-$ SLJ-trained}} \\
judge & $16.0\ [13.1,\,18.9]$ & $2.1\ [0.2,\,4.2]$ & $0.7\ [-1.2,\,2.6]$ \\
judge $+$ gate & $56.3\ [52.6,\,59.9]$ & $37.3\ [33.0,\,41.8]$ & $30.6\ [25.9,\,35.2]$ \\
execution (hardened) & $22.3\ [18.8,\,26.1]$ & $18.1\ [14.6,\,21.9]$ & $15.9\ [12.1,\,20.0]$ \\
\bottomrule
\end{tabular}
\end{table}

\section{Simple checks and controlled syntax errors}
\label{app:signals}

\paragraph{Simple checks.} We compared three cheap features for distinguishing damaged-run samples from base-model samples. The score is the probability that a randomly chosen damaged-run sample ranks above a base-model sample, with ties counted as one half ($20{,}000$ samples each, pooled over prompts). A score of $50\%$ means no separation. For the binary parse flag, ties are common, so this score is larger than the probability of a strict ranking. The parse failure flag reaches $83\%$ for the SLJ-trained checkpoint and $81\%$ for the CE-trained one. Output length reaches $67\%$ and $63\%$, and the share of the response that is code reaches $67\%$ and $63\%$ (damaged outputs contain a larger share of code). Within a damaged run, length separates non-parsing from parsing samples with probability $67\%$. The parse flag provides the strongest separation, although output length also carries some information. None of these features measures correctness.

\paragraph{Controlled syntax errors.} The damage in the trained checkpoints is severe, so we asked how the instruments respond to milder damage. We took the $13{,}472$ base-model samples ($32$ per problem, $421$ problems) and inserted one surplus bracket into a nested, seeded fraction of the samples (nominal doses $0$, $0.1$, $0.3$, $0.6$ and $1$, realised corrupted shares $0$, $10.2\%$, $30.2\%$, $59.6\%$ and $99.5\%$). We could corrupt $13{,}404$ samples, while $68$ were already invalid. We then re-judged all of them with the strong judge. Re-judging all $13{,}472$ original responses reproduces the cached verdict for $99.9\%$ of them. On the corrupted samples it accepts $60.2\%$, against $84.1\%$ for the same samples before corruption, so it notices some of the damage and still accepts most of it. 

The effect on Pass@$32$ is small: the judge scores $96.2\%$ at dose $0$ and $96.0\%$ at dose $0.6$, the gated judge $96.0\%$ and $94.8\%$, and hardened execution $42.8\%$ and $41.1\%$. At Pass@$1$ the judge falls from $84.0\%$ to $69.2\%$ over the same range, and the gate from $83.7\%$ to $33.6\%$. At dose $1$ the judge still scores $91.9\%$ at Pass@$32$, and the gate and execution both score $0\%$. 

This experiment varies the share of answers with syntax errors, while keeping the type of error fixed: every corrupted sample carries one surplus bracket, whereas real damage can be larger. It also cannot say anything about the cause of the damage in the trained checkpoints. For reference, the parse rate of the full Qwen base replay is $99.6\%$, while the uncorrupted samples in this experiment parse at $99.5\%$.

\section{Judge inputs, verdict matching and a second judge}
\label{app:judges}

\paragraph{Text outside the code.} The judge reads the whole response, while the parse gate reads only the extracted code. Almost all responses are code alone: text of more than $40$ characters outside all fenced code blocks occurs in $51$ of $53{,}888$ base samples and in $146$ and $222$ samples of the CE- and SLJ-trained checkpoints. Where it occurs, the judge accepts less often, not more. For non-parsing samples the acceptance rate is $60.7\%$ ($[56.8\%,\,64.7\%]$) without such text and $31.0\%$ ($[19.3\%,\,40.2\%]$) with it for the SLJ-trained checkpoint, and $59.4\%$ against $35.0\%$ ($[19.8\%,\,47.8\%]$) for the CE-trained one. Text outside the code therefore does not explain why the judge accepts non-parsing programs. The groups with text are small ($216$ and $137$ non-parsing samples from $65$ and $58$ prompts), and the intervals resample only the prompts that contribute to a group, so they are wide and conditional.

\paragraph{Judge prompt, input length and pairing.} The judge prompt asks for a single character, \texttt{1} if the candidate answer is correct and \texttt{0} otherwise, and shows the problem, the reference solution as ``gold answer'' and the full model response. The template has not changed in the code since before the study's runs (SHA-256 \texttt{5329b0bf...962c1}), and both judges use it. Its full text, with the three fields filled in per call, is:
\begin{small}
\begin{verbatim}
You are an expert grader. Your job is to decide whether a candidate
answer is CORRECT or INCORRECT for the given problem.

Output exactly one character: '1' if the candidate answer is correct,
or '0' if it is incorrect. Do not output anything else — no reasoning,
no punctuation, no whitespace.

PROBLEM:
{prompt}

GOLD ANSWER:
{gold}

CANDIDATE ANSWER:
{response}

YOUR VERDICT (1 or 0):
\end{verbatim}
\end{small}
For the Qwen base model's second-run answers, we re-judged $1{,}280$ cached
verdicts with Qwen2.5-72B-Instruct at temperature $0$. We sampled $80$
questions from each benchmark and $8$ answers per question. None of the
requests failed or exceeded the $4096$-token context limit (the longest had
$1{,}187$ input tokens), and every reply was a valid \texttt{0} or \texttt{1}.
The new verdict matched the cached verdict for $1{,}276$ of $1{,}280$ items
($99.69\%$). Agreement was $88.98\%$ when the cached sample index was shifted
by one, and $76.09\%$ when the cached verdict came from a randomly selected
other question. These checks support correct pairing in the sampled Qwen
base-model answers. They do not prove that the original requests were
identical, and they do not test pairing in the trained runs. The baselines
use one shift and one other-question draw per item. In one disagreement the
cached verdict was $0$ and the new verdict $1$. In the other three, the
cached verdict was $1$ and the new verdict $0$. The Qwen-only counts are
saved in \texttt{tables/judge\_recheck\_qwen\_summary.json} in the code release.

\paragraph{A weaker judge on the damaged checkpoints.} We also had the weak judge (Qwen2.5-7B-Instruct, same prompt) grade all $53{,}888$ second-run samples of each damaged checkpoint. It accepts fewer samples overall than the strong judge and contrasts parsing with non-parsing samples more sharply: it accepts $28.0\%$ of the non-parsing samples of the CE-trained checkpoint and $30.4\%$ of those of the SLJ-trained one, against $80.4\%$ and $81.6\%$ of their parsing samples (the strong judge accepts $59.3\%$ and $60.6\%$ of the non-parsing samples). Its verdicts agree with the strong judge's on $72.6\%$ and $72.1\%$ of samples, and it accepts $1{,}962$ (CE) and $1{,}916$ (SLJ) samples that the strong judge rejects. 

Its Pass@$128$ remains $85.7\%$ (CE) and $84.8\%$ (SLJ), against hardened execution at $30.2\%$ and $28.3\%$ and the strong judge at $95.2\%$ and $96.4\%$. At Pass@$32$ it is $80.0\%$ and $80.0\%$ against execution $26.3\%$ and $24.7\%$. Applying the parse gate to the weak judge lowers Pass@$128$ to $67.5\%$ and $66.5\%$. For this judge on these outputs, the gap to execution narrows but does not close (weak minus execution at Pass@$128$: $55.6$ pp, 95\% interval $[50.6,\,60.3]$ pp for CE, and $56.5$ pp, $[51.5,\,61.3]$ pp for SLJ). 

Compared with the first run's weak-judge Pass@$32$ ($96.2\%$ for CE and $96.0\%$ for SLJ), the weak judge's score falls by about $16$ pp, far more than the strong judge's $3$ pp, so the insensitivity reported in the main text belongs to the strong judge rather than to every judge. This cross-run comparison carries the same drift caveat as the strong judge's. We have no weak-judge verdicts for the base-model replay samples, so we report no base-minus-damaged separation for this judge, and the verdict-to-sample pairing is assumed as for the strong judge. Both judges belong to the Qwen2.5 family, so this tests a change of judge size within one family, not other judge families.

\section{Generation settings and artifact matching}
\label{app:settings}
Table~\ref{tab:settings} lists what the generation script records and what it does not. The replay (k=128) used the repository's evaluation script with vLLM, loading the tokenizer from each checkpoint directory and applying that tokenizer's chat template with the system prompt and the problem as user turn. The archived run (k=32) came from the same script family. Its saved summaries record the sampling parameters (temperature $1.0$, top-$p$ $0.95$, $1024$ new tokens), which match the replay. Tokenizer and chat template, context length, stop and EOS handling and checkpoint identity were not recorded, so we cannot compare the two runs on those. The replay's system prompt, the same text the training script uses, is:
\begin{small}
\begin{verbatim}
You are a Python expert. Write a Python solution. Wrap your final
solution in a single ```python ... ``` code block. Be concise; do not
write anything after the code block.
\end{verbatim}
\end{small}
The first-run generations still exist, but we did not re-judge them under the current judge configuration, so the cross-run comparison keeps the temporal caveat of the main text. For future evaluations we recommend recording the tokenizer and template version, stop tokens, software versions, a checksum of each checkpoint and the parse rate of every run. The missing records limit this study and leave the cause of the discrepancy unresolved. We also found no checksums or parameter statistics for the trained checkpoints, recorded at training or later, so we cannot show that the replayed weights are the weights that produced the archived samples.

\begin{table}[h]
\centering
\caption{Generation settings. ``Replay'' is the $k=128$ run examined here, while ``archived'' is the earlier $k=32$ run.}
\label{tab:settings}
\small
\begin{tabular}{lll}
\toprule
setting & replay ($k=128$) & archived ($k=32$) \\
\midrule
sampling & temperature $1.0$, top-$p$ $0.95$ & temperature $1.0$, top-$p$ $0.95$ \\
max new tokens & $1024$ & $1024$ \\
max model length & $4096$ & not recorded \\
tokenizer and chat template & loaded from each checkpoint directory & not recorded \\
stop tokens, EOS handling & vLLM defaults, no custom stop strings & not recorded \\
checkpoint checksums & none found & none found \\
\bottomrule
\end{tabular}
\end{table}

\section{Additional evaluation statistics}
\label{app:additionalstats}

The following tables support the comparisons between runs, the effect of
the syntax gate, and the two additional checks discussed in
Sections~\ref{sec:judge}--\ref{sec:limitations}.
The comparisons between checkpoints at different attempt budgets are in
Tables~\ref{tab:figurevalues} and~\ref{tab:paired}.

\subsection{Comparison between the two runs}

\begin{table}[H]
\centering
\caption{Comparison of the two runs using the Qwen-2.5-72B-Instruct judge at
Pass@$32$. The right-hand columns show the percentage of answers that parse.
Changes are in percentage points (pp) and use scores before rounding.
Approximate $95\%$ intervals for the judge-score change are
$[-2.41, +2.69]$ pp for the base, $[-6.01, -0.04]$ pp for CE-trained and
$[-5.38, -0.07]$ pp for SLJ-trained. These intervals are unpaired because
first-run per-question judge verdicts were not retained.}
\label{tab:temporal}
\small
\begin{tabular}{lccc|ccc}
\toprule
& \multicolumn{3}{c|}{judge Pass@$32$} & \multicolumn{3}{c}{samples parsing} \\
checkpoint & first & second & change (pp) & first & second & change (pp) \\
\midrule
\texttt{Qwen-2.5-7B} base & $95.96\%$ & $96.10\%$ & $+0.14$ & $99.61\%$ & $99.60\%$ & $-0.01$ \\
CE-trained                & $95.72\%$ & $92.70\%$ & $-3.03$ & $99.58\%$ & $36.96\%$ & $-62.62$ \\
SLJ-trained               & $96.67\%$ & $93.95\%$ & $-2.73$ & $99.58\%$ & $33.10\%$ & $-66.48$ \\
\bottomrule
\end{tabular}
\end{table}

For these between-run intervals, let $p_1$ be the first-run judge
Pass@$32$ score and $v_i$ the second-run Pass@$32$ estimate for question
$i$, calculated from its $128$ judge verdicts. With $N=421$, the estimated
change is $\Delta=\bar v-p_1$. We use the normal approximation
\[
\Delta \;\pm\; 1.96\sqrt{\frac{p_1(1-p_1)}{N}
                 +\frac{s_v^2}{N}},
\]
where $s_v^2$ is the sample variance of the $421$ values $v_i$, using
denominator $N-1$. At $k=32$, each first-run question either has an approved
answer among its $32$ samples or does not, which gives the first variance
term. The second term measures variation in the second-run question-level
estimates. We cannot estimate the covariance between runs without the
first-run question-level verdicts, so the formula treats the runs as
unpaired. All calculations use unrounded scores. The retained inputs,
standard errors and intervals are listed in
\texttt{tables/temporal\_ci\_recomputed.json} in the code release.

\subsection{Effect of the syntax gate}

\begin{table}[H]
\centering
\caption{Effect of adding the syntax gate to judge-based Pass@$1$ on the
second-run answers. Scores are percentages. The drop is judge score minus
gated score, in percentage points (pp), computed before rounding. Intervals
are paired $95\%$ bootstrap intervals over $421$ questions with $5000$
resamples. Across the three checkpoints, none of the $70{,}239$ invalid
answers passed the hardened execution tests.}
\label{tab:gatecost}
\small
\setlength{\tabcolsep}{4pt}
\begin{tabular}{lcccc}
\toprule
checkpoint & Judge (\%) & Judge + gate (\%) & Drop (pp) & Interval (pp) \\
\midrule
base & $84.12$ & $83.92$ & $0.19$ & $[0.11, 0.28]$ \\
CE-trained & $67.45$ & $30.05$ & $37.40$ & $[34.11, 40.96]$ \\
SLJ-trained & $68.10$ & $27.59$ & $40.51$ & $[37.31, 43.82]$ \\
\bottomrule
\end{tabular}
\end{table}

\subsection{Controlled bracket errors}

\begin{table}[H]
\centering
\caption{Controlled bracket errors added to saved base-model answers.
The first column gives the target share of answers to modify. Actual shares
reflect the seeded selection, with already-invalid answers left unchanged.
Parsing and judge Pass@$32$ scores are percentages.
The change in judge score is relative to the unmodified answers, in
percentage points (pp). The procedure is described in
Appendix~\ref{app:signals}.}
\label{tab:corruption}
\small
\setlength{\tabcolsep}{4pt}
\begin{tabular}{rrrrr}
\toprule
\shortstack{Target share\\(\%)} & \shortstack{Modified\\(\%)} &
\shortstack{Parsing\\(\%)} & \shortstack{Judge Pass@$32$\\(\%)} &
\shortstack{Judge change\\(pp)} \\
\midrule
$0$ & $0.00$ & $99.50$ & $96.20$ & $+0.00$ \\
$10$ & $10.17$ & $89.33$ & $96.20$ & $+0.00$ \\
$30$ & $30.17$ & $69.33$ & $95.96$ & $-0.24$ \\
$60$ & $59.57$ & $39.93$ & $95.96$ & $-0.24$ \\
$100$ & $99.50$ & $0.00$ & $91.92$ & $-4.28$ \\
\bottomrule
\end{tabular}
\end{table}

\subsection{Supplying the expected MBPP function names}

\begin{table}[H]
\centering
\caption{Execution Pass@$128$ on the same $257$ MBPP questions, with and
without the expected function name in the prompt. Values are percentages
under the hardened execution tests. The named-prompt runs generated new
answers, with $128$ attempts per question, and do not replace the original
evaluation results. Appendix~\ref{app:hardened} explains the naming issue.}
\label{tab:mbppnames}
\small
\begin{tabular}{lcc}
\toprule
base model & Original prompts (\%) & Function name supplied (\%) \\
\midrule
Qwen-2.5-7B & $12.45$ & $79.38$ \\
\bottomrule
\end{tabular}
\end{table}

%% file: references.bib
@article{shao2024grpo,
  title   = {{DeepSeekMath}: Pushing the Limits of Mathematical Reasoning in Open Language Models},
  author  = {Shao, Zhihong and Wang, Peiyi and Zhu, Qihao and Xu, Runxin and Song, Junxiao and others},
  journal = {arXiv preprint arXiv:2402.03300},
  year    = {2024},
}

@article{guo2025deepseekr1,
  title   = {{DeepSeek-R1} Incentivizes Reasoning in {LLMs} through Reinforcement Learning},
  author  = {Guo, Daya and Yang, Dejian and Zhang, Haowei and Song, Junxiao and Wang, Peiyi and Zhu, Qihao and Xu, Runxin and Zhang, Ruoyu and Ma, Shirong and Bi, Xiao and others},
  journal = {Nature},
  volume  = {645},
  number  = {8081},
  pages   = {633--638},
  year    = {2025},
  doi     = {10.1038/s41586-025-09422-z},
}

@article{qwen2024qwen25,
  title   = {{Qwen2.5} Technical Report},
  author  = {{Qwen Team}},
  journal = {arXiv preprint arXiv:2412.15115},
  year    = {2024},
}

@article{austin2021mbpp,
  title   = {Program Synthesis with Large Language Models},
  author  = {Austin, Jacob and Odena, Augustus and Nye, Maxwell and others},
  journal = {arXiv preprint arXiv:2108.07732},
  year    = {2021},
}

@article{chen2021humaneval,
  title   = {Evaluating Large Language Models Trained on Code},
  author  = {Chen, Mark and Tworek, Jerry and Jun, Heewoo and others},
  journal = {arXiv preprint arXiv:2107.03374},
  year    = {2021},
}

@inproceedings{kwon2023vllm,
  title     = {Efficient Memory Management for Large Language Model Serving with {PagedAttention}},
  author    = {Kwon, Woosuk and Li, Zhuohan and Zhuang, Siyuan and Sheng, Ying and Zheng, Lianmin and Yu, Cody Hao and Gonzalez, Joseph E. and Zhang, Hao and Stoica, Ion},
  booktitle = {Proceedings of the 29th Symposium on Operating Systems Principles (SOSP)},
  year      = {2023},
}

@inproceedings{gao2023overoptimization,
  title     = {Scaling Laws for Reward Model Overoptimization},
  author    = {Gao, Leo and Schulman, John and Hilton, Jacob},
  booktitle = {International Conference on Machine Learning (ICML)},
  year      = {2023},
}

@inproceedings{hochlehnert2025soberlook,
  title     = {A Sober Look at Progress in Language Model Reasoning: Pitfalls and Paths to Reproducibility},
  author    = {Hochlehnert, Andreas and Bhatnagar, Hardik and Udandarao, Vishaal and Albanie, Samuel and Prabhu, Ameya and Bethge, Matthias},
  booktitle = {Conference on Language Modeling (COLM)},
  year      = {2025},
}

@inproceedings{panickssery2024selfpref,
  title     = {{LLM} Evaluators Recognize and Favor Their Own Generations},
  author    = {Panickssery, Arjun and Bowman, Samuel R. and Feng, Shi},
  booktitle = {Advances in Neural Information Processing Systems (NeurIPS)},
  year      = {2024},
}

@inproceedings{shao2025spuriousrewards,
  title     = {Spurious Rewards: Rethinking Training Signals in {RLVR}},
  author    = {Shao, Rulin and Li, Shuyue Stella and Xin, Rui and Geng, Scott and Wang, Yiping and Oh, Sewoong and Du, Simon Shaolei and Lambert, Nathan and Min, Sewon and Krishna, Ranjay and Tsvetkov, Yulia and Hajishirzi, Hannaneh and Koh, Pang Wei and Zettlemoyer, Luke},
  booktitle = {Proceedings of the 43rd International Conference on Machine Learning (ICML)},
  series    = {Proceedings of Machine Learning Research},
  volume    = {306},
  pages     = {110167--110207},
  year      = {2026},
}

@article{helff2026gamingverifiers,
  title   = {{LLMs} Gaming Verifiers: {RLVR} can Lead to Reward Hacking},
  author  = {Helff, Lukas and Delfosse, Quentin and Steinmann, David and H{\"a}rle, Ruben and Shindo, Hikaru and Schramowski, Patrick and Stammer, Wolfgang and Kersting, Kristian and Friedrich, Felix},
  journal = {arXiv preprint arXiv:2604.15149},
  year    = {2026},
}

@article{wu2025hiddencosts,
  title   = {Position: The Hidden Costs and Measurement Gaps of Reinforcement Learning with Verifiable Rewards},
  author  = {Wu, Fang and Tu, Aaron and Xuan, Weihao and Qi, Heli and Huang, Xu and others},
  journal = {arXiv preprint arXiv:2509.21882},
  year    = {2025},
}

@inproceedings{lambert2024tulu3,
  title     = {{T{\"u}lu 3}: Pushing Frontiers in Open Language Model Post-Training},
  author    = {Lambert, Nathan and Morrison, Jacob and Pyatkin, Valentina and Huang, Shengyi and Ivison, Hamish and Brahman, Faeze and Miranda, Lester James V. and Liu, Alisa and Dziri, Nouha and Lyu, Xinxi and Gu, Yuling and Malik, Saumya and Graf, Victoria and Hwang, Jena D. and Yang, Jiangjiang and Le Bras, Ronan and Tafjord, Oyvind and Wilhelm, Chris and Soldaini, Luca and Smith, Noah A. and Wang, Yizhong and Dasigi, Pradeep and Hajishirzi, Hannaneh},
  booktitle = {Conference on Language Modeling (COLM)},
  year      = {2025},
  note      = {Extended preprint at arXiv:2411.15124},
}

@inproceedings{zhao2025codejudgeeval,
  title     = {{CodeJudge-Eval}: Can Large Language Models be Good Judges in
               Code Understanding?},
  author    = {Zhao, Yuwei and Luo, Ziyang and Tian, Yuchen and Lin, Hongzhan
               and Yan, Weixiang and Li, Annan and Ma, Jing},
  booktitle = {Proceedings of the 31st International Conference on
               Computational Linguistics (COLING)},
  pages     = {73--95},
  year      = {2025},
}

@inproceedings{jiang2025codejudgebench,
  title     = {{CodeJudgeBench}: Benchmarking {LLM}-as-a-Judge for Coding Tasks},
  author    = {Jiang, Hongchao and Chen, Yiming and Cao, Yushi and Lee, Hung-yi
               and Tan, Robby T.},
  booktitle = {Proceedings of the 64th Annual Meeting of the Association for Computational Linguistics (Volume 1: Long Papers)},
  pages     = {19416--19448},
  year      = {2026},
  doi       = {10.18653/v1/2026.acl-long.888},
}

@article{wang2025axiom,
  title   = {Benchmarking Code {LLM}-as-a-Judge via Constrained Perturbation and Calibration},
  author  = {Wang, Ruiqi and Wang, Xinchen and Gao, Cuiyun and Chong, Chun Yong
             and Xia, Xin and Liao, Qing},
  journal = {arXiv preprint arXiv:2512.20159v2},
  year    = {2026},
}

@inproceedings{zheng2023judging,
  title     = {Judging {LLM}-as-a-Judge with {MT-Bench} and {Chatbot Arena}},
  author    = {Zheng, Lianmin and Chiang, Wei-Lin and Sheng, Ying and Zhuang, Siyuan and Wu, Zhanghao and Zhuang, Yonghao and Lin, Zi and Li, Zhuohan and Li, Dacheng and Xing, Eric P. and Zhang, Hao and Gonzalez, Joseph E. and Stoica, Ion},
  booktitle = {Advances in Neural Information Processing Systems (NeurIPS), Datasets and Benchmarks Track},
  year      = {2023},
}

@inproceedings{tong2024codejudge,
  title     = {{CodeJudge}: Evaluating Code Generation with Large Language Models},
  author    = {Tong, Weixi and Zhang, Tianyi},
  booktitle = {Proceedings of the 2024 Conference on Empirical Methods in Natural Language Processing (EMNLP)},
  pages     = {20032--20051},
  year      = {2024},
  doi       = {10.18653/v1/2024.emnlp-main.1118},
}

@inproceedings{liu2023evalplus,
  title     = {Is Your Code Generated by {ChatGPT} Really Correct? Rigorous Evaluation of Large Language Models for Code Generation},
  author    = {Liu, Jiawei and Xia, Chunqiu Steven and Wang, Yuyao and Zhang, Lingming},
  booktitle = {Advances in Neural Information Processing Systems (NeurIPS)},
  year      = {2023},
}

@inproceedings{yuan2024selfrewarding,
  title     = {Self-Rewarding Language Models},
  author    = {Yuan, Weizhe and Pang, Richard Yuanzhe and Cho, Kyunghyun and Li, Xian and Sukhbaatar, Sainbayar and Xu, Jing and Weston, Jason},
  booktitle = {Proceedings of the 41st International Conference on Machine Learning (ICML)},
  series    = {Proceedings of Machine Learning Research},
  volume    = {235},
  pages     = {57905--57923},
  year      = {2024},
}

@article{romera2024funsearch,
  title   = {Mathematical Discoveries from Program Search with Large Language Models},
  author  = {Romera-Paredes, Bernardino and Barekatain, Mohammadamin and Novikov, Alexander and Balog, Matej and Kumar, M. Pawan and Dupont, Emilien and Ruiz, Francisco J. R. and Ellenberg, Jordan S. and Wang, Pengming and Fawzi, Omar and Kohli, Pushmeet and Fawzi, Alhussein},
  journal = {Nature},
  volume  = {625},
  number  = {7995},
  pages   = {468--475},
  year    = {2024},
  doi     = {10.1038/s41586-023-06924-6},
}

@article{novikov2025alphaevolve,
  title   = {{AlphaEvolve}: A Coding Agent for Scientific and Algorithmic Discovery},
  author  = {Novikov, Alexander and V{\~u}, Ng{\^a}n and Eisenberger, Marvin and Dupont, Emilien and Huang, Po-Sen and Wagner, Adam Zsolt and Shirobokov, Sergey and Kozlovskii, Borislav and Ruiz, Francisco J. R. and Mehrabian, Abbas and Kumar, M. Pawan and See, Abigail and Chaudhuri, Swarat and Holland, George and Davies, Alex and Nowozin, Sebastian and Kohli, Pushmeet and Balog, Matej},
  journal = {arXiv preprint arXiv:2506.13131},
  year    = {2025},
}

@inproceedings{le2022coderl,
  title     = {{CodeRL}: Mastering Code Generation through Pretrained Models and Deep Reinforcement Learning},
  author    = {Le, Hung and Wang, Yue and Gotmare, Akhilesh Deepak and Savarese, Silvio and Hoi, Steven C. H.},
  booktitle = {Advances in Neural Information Processing Systems (NeurIPS)},
  year      = {2022},
}

@article{shojaee2023ppocoder,
  title   = {Execution-based Code Generation using Deep Reinforcement Learning},
  author  = {Shojaee, Parshin and Jain, Aneesh and Tipirneni, Sindhu and Reddy, Chandan K.},
  journal = {Transactions on Machine Learning Research},
  year    = {2023},
}

@inproceedings{yuan2025nondeterminism,
  title     = {Understanding and Mitigating Numerical Sources of Nondeterminism in {LLM} Inference},
  author    = {Yuan, Jiayi and Li, Hao and Ding, Xinheng and Xie, Wenya and Li, Yu-Jhe and Zhao, Wentian and Wan, Kun and Shi, Jing and Hu, Xia and Liu, Zirui},
  booktitle = {Advances in Neural Information Processing Systems (NeurIPS)},
  year      = {2025},
}
